\pdfoutput=1

\documentclass[12pt, a4paper]{iopart}

\usepackage{hyperref}
\usepackage{epsfig}
\usepackage{graphicx,epsf}
\usepackage{color}
\usepackage{bm}
\usepackage{iopams}

\eqnobysec

\def\be{\begin{equation}}
\def\ee{\end{equation}}

\def\bea{\begin{eqnarray}}
\def\eea{\end{eqnarray}}

\def\R{{{\cal{R}}}}

\def\J{{{\cal{J}}}}
\def\H{{\cal H}}

\def\cs2{c_{\rm{s}}^2}

\def\U0{{\bar U_0}}

\def\M{{\cal{M}}}

\def\bi{\begin{itemize}}
\def\ei{\end{itemize}}

\newcommand\eq[1]{Eq.~(\ref{#1})}

\usepackage{epsfig}
\usepackage{graphicx,epsf}
\usepackage{color}
\usepackage{bm}
\usepackage{psfrag}

\def\be{\begin{equation}}
\def\ee{\end{equation}}
\def\beb{\begin{equation*}}
\def\eeb{\end{equation*}}
\def\bea{\begin{eqnarray}}
\def\eea{\end{eqnarray}}
\def\beab{\begin{eqnarray*}}
\def\eeab{\end{eqnarray*}}
\def\nn{\nonumber}

\def\R{{{\cal{R}}}}
\def\Ps{{{\cal{P}}}}

\def\H{{\cal H}}

\def\cs2{c_{\rm{s}}^2}

\def \beg {\begin{enumerate}}
\def \en {\end{enumerate}}

\def\drho{{\delta\rho}}
\def\dP{{\delta P}}

\def\cs{c_{\rm{s}}^2}

\def\Mt{{\cal M}}
\def\Et{{\cal E}}

 \def\MII{{\cal M_{\rm{2}}}}
 
 \def\EI{{\cal E_{\rm{1}}}}
 \def\EII{{\cal E_{\rm{2}}}}

\begin{document}

\today

\title{Effects of non-linearities on magnetic field generation}

\author{Ellie Nalson$^1$, Adam J.~Christopherson$^2$, Karim A.~Malik$^1$}

\address{$^1$Astronomy Unit, School of Physics and Astronomy, Queen Mary University of London,
Mile End Road, London, E1 4NS, UK}
\address{$^2$School of Physics and Astronomy, University of Nottingham, University Park,
Nottingham, NG7 2RD, UK}

\begin{abstract} 
Magnetic fields are present on all scales in the Universe. While we
understand the processes which amplify the fields fairly well, we do
not have a ``natural'' mechanism to generate the small initial seed
fields.  By using fully relativistic cosmological perturbation theory
and going beyond the usual confines of linear theory we show
analytically how magnetic fields are generated.
This is the first analytical calculation of the magnetic field at second order, using gauge-invariant cosmological perturbation theory, and including \emph{all} the source terms. To this end, we have rederived the full set of governing equations independently.
Our results suggest that magnetic fields of the order of $10^{-30}-10^{-27}$ G can be 
generated (although this depends on the small scale cut-off of the integral), which is largely in agreement with previous results that relied upon numerical calculations.
These fields are likely too small to act as the
primordial seed fields for dynamo mechanisms.
\end{abstract}


\maketitle

\section{Introduction}
\label{sec:intro}

Magnetic fields are prevalent everywhere in the universe, from the
small scales in our solar system to larger, intergalactic scales
\cite{Brown:2006wv}. These fields are relatively strong on planetary
scales, of the order of a few Gauss, and have a coherence length of a
few thousand kilometres, but become weaker as the scales, and the
fields' coherence length, increase. On galactic scales, magnetic
fields are observed with a coherence length of a few kiloparsecs and a
strength of around $1\mu {\rm G}$ \cite{Widrow:2002ud, Kulsrud:2007an,
  Kronberg:1993vk, Carilli:2001hj}, while on galaxy cluster scales
similar strength magnetic fields are found with larger coherence
lengths, of a few megaparsecs \cite{Kronberg:2007dy, Bernet:2008qp,
  Wolfe:2008nk}. Recently there have been some exciting observations
showing the existence of inter-cluster magnetic fields within voids,
with strengths between $10^{-17} - 10^{-14}{\rm G}$
\cite{Tavecchio:2010ja, Ando:2010rb, Neronov:1900zz, Tavecchio:2010mk,
  Essey:2010nd}.

Despite their importance, surprisingly little is known about the
origin of the magnetic fields in our universe. While astrophysical
mechanisms could account for some of the fields on smaller scales, the
fact that magnetic fields appear to exist also on very large scales,
and at large redshift suggest that they are cosmological in origin.

The presence of magnetic fields in present-day galaxies can perhaps be
explained by the amplification of small seed fields by either the
dynamo mechanism \cite{moffatbook, Kulsrud:1992rk, han, Grasso:2000wj,
  Kulsrud:1996km}, or by the adiabatic compression of a previously
magnetised cloud \cite{Grasso:2000wj, King:2005xh}.
The dynamo mechanisms require a seed field with strength between
$10^{-12}$G and $10^{-30}$G in order to satisfy observational
constraints, while amplification by adiabatic compression is not as
efficient as the dynamo, and requires a larger seed field of at least
$10^{-20}$G.

While both these mechanisms can explain the magnetic fields observed
on galactic and possible cluster scales, they face difficulties with
those observed at high redshift and even more difficulties with the
intergalactic fields. Additionally, the question remains: what is the
origin of the seed magnetic field?
There are many explanations for the origin of the seed magnetic
fields, each with its own problem. Astrophysical processes after
recombination and battery-type effects, such as the Biermann-battery
\cite{biermann, daly:1990, subramanian1994thermal, Gnedin:2000ax,
  Kulsrud:1996km, Davies:1999bk, Giovannini:2003yn} or supernova
batteries \cite{Hanayama:2005hd, Miranda:1998ne}, are one possible
solution. However, although these are strong enough to seed dynamos,
these processes only work on galactic scales and so cannot source
magnetic fields on cluster or intergalactic scales. Therefore, we
suppose that magnetic fields were formed at an earlier time than when
these processes are at work.

The generation of magnetic fields in the very early universe has been
the focus of many studies in the literature, for example
Refs.~\cite{Grasso:2000wj, Giovannini:2003yn, Turner:1987bw,
  Tornkvist:2000js, Dimopoulos:2001wx, Prokopec:2004au, Bamba:2004cu,
  Bassett:2000aw, Marklund:2000zs}. There are many such methods, all which have their own flaws, and sustaining magnetic fields in the early universe proves difficult. Most of these methods fall into the following categories: quantum-mechanically generated fields during inflation, field generation through phase transitions such as electroweak symmetry breaking, magnetic fields generated during (p)reheating.
  
Additionally, magnetic fields could have been created by vorticity, in a
process first investigated by Harrison \cite{harrison}. Here, the
fields could be created continuously in a period between lepton
decoupling and recombination by vorticity naturally occurring in
higher order perturbation theory \cite{Matarrese:2004kq, Gopal:2004ut,
  Takahashi:2005nd, Betschart:2003bn}. This process will be the focus of the present
work.

In addition to acting as a seed for the dynamo mechanisms, the
primordial magnetic field must satisfy other observational
constraints. These come from nucleosynthesis, gravitational waves and
various CMB observables such as the magnetised Sunyaev-Zeldovich effect,
Faraday rotations and cosmological perturbations
\cite{Durrer:2013pga}.

Magnetic fields can have post-recombination effects which put an upper
bound on their strength. For instance magnetic fields can affect the
thermal and chemical evolution of the Intergalactic Medium (IGM)
during dark ages. The dissipation of a small fraction of the magnetic
field energy increases the temperature, enhancing the ionisation
fraction of the IGM and leading to larger molecule
abundances. Magnetic fields also affect the formation of the first
stars through changing their mass scale due to the magnetic Jeans mass
dominating over the thermal Jeans mass. The magnetic fields would also
impact upon the epoch of reionisation which could potentially be
detectable through future 21cm experiments \cite{Tashiro:2006uv,
  Schleicher:2008hc}.\\

In this paper we consider magnetic field generation in cosmological
perturbation theory, working up to second order. There have been fully
numerical studies reported in the literature, focusing on specific
terms in the evolution equations, such as in
Refs.~\cite{Ichiki:2007hu, Maeda:2011uq}.  The full set of governing
equations has been solved numerically in Ref.~\cite{Fenu:2010kh}.
Here we present the first complete study using analytical techniques
throughout.  First, we derive the governing equations for the electric
and magnetic field up to second order in metric perturbation
theory. We then compute the power spectrum of the resultant magnetic
field, comparing to previous results where appropriate. This is the
first analytical calculation of the magnetic field at second order
that has included all the source terms -- where previous analytical
calculations have been performed, they omitted the particularly tricky
part of the source term (e.g. Ref.~\cite{Takahashi:2005nd}). As will be shown, the magnetic field is generated, in part, by non-adiabatic pressure perturbations. In this work, we consider two sources of non-adiabatic pressure: the isocurvature perturbations left over from inflation, and imprinted in the CMB, and the relative non-adiabatic pressure arising from the multi-component nature of the cosmic fluid.
Our
analytical calculations largely agree with previous results and the
magnetic field that is generated at second order in perturbation
theory is likely too weak to act as the primordial seed field for
later, astrophysical battery-type mechanisms.

We can get an idea on how the, at first glance, different generation
mechanisms listed above are related by considering the ``naive'' magnetic
field constraint equation (see Eq.~(\ref{Maxwell1}) below for the
``full'' equation)
\be
\label{naive_mag}
\Mt^k_{;k}\simeq\omega^i\, \Et_i\,. 
\ee
We can see in the above equation the close relation of the magnetic
field, $\mathcal{M}^k$ defined in Eq.~(\ref{eq:defM}), to vorticity, $\omega^i$ defined in Eq.~(\ref{eq:vortdef}), and hence the generation of magnetic fields is very similar and related to the generation of vorticity. In the above, $\mathcal{E}^i$ is the electric field defined in Eq.~(\ref{eq:defE}).

There are now several possibilities to use \eq{naive_mag} to generate
magnetic fields. One possibility is to generate vorticity explicitly,
e.g.~by introducing shocks into the system as in
Ref.~\cite{Ryu16052008}. Alternatively, we can get vorticity by requiring or directly
prescribing the velocity field to have rotational components, or use
the velocity difference in the fluids present, as in the classic paper
by Harrison \cite{harrison}.

Another possibility is to take the time derivative of \eq{naive_mag},
and we immediately get the classical ``Biermann battery'', since
$\dot\omega^i$ is sourced by the gradients of energy density and the
non-adiabatic pressure perturbation \cite{biermann}. We follow a very similar route in this work, allowing for
gradients in the energy density and the non-adiabatic pressure or
entropy perturbation, however, using cosmological perturbation theory
which allows us to study the problem in full generality.


The paper is structured as follows: in the next section we introduce
magnetic fields in a cosmological setting, providing a brief
introduction to cosmological perturbation theory, magnetic fields and
the Maxwell equations, followed by a derivation of the
evolution equations of the magnetic and electric fields up to second order in
perturbation theory. In Section~\ref{sec:results}, we solve the
governing equations and present our results. We summarise our findings
in Section~\ref{sec:dis} and conclude with a discussion of potential
future work.

\section{Magnetic fields in cosmology}
\label{sec:basics}

First, we introduce the formalism and equations governing a cosmological system including electromagnetism. For more detail, we direct the interested reader to, e.g., Ref.~\cite{Barrow:2006ch}, although we stress that in this article we use metric cosmological perturbation theory throughout.

\subsection{Cosmological perturbations}

In this paper we consider perturbations to a FLRW spacetime and work in the uniform curvature gauge, neglecting tensor perturbations,\footnote{Although vectors and tensors couple at higher order, the vector modes after inflation are negligible, and the gravitational wave contribution is small. Since we are interested in the magnetic field from scalar perturbations, we neflect tensors and, later, vectors in this work.} in which the line element takes the form \cite{Bardeen:1980kt,ks}
\be 
ds^2=a^2(\eta)\Big[-(1+2\phi)d\eta ^2+2aB_{i}dx^id\eta +\delta_{ij}dx^idx^j\Big]\,.
\ee
Here $a(\eta)$ is the scale factor, $\eta$ denotes the conformal time coordinate, $\phi$ is the lapse function and $B^i$ is the shear. Throughout this paper, Greek indices ($\kappa, \lambda, \mu, \ldots$) denote  full spacetime indices, Latin letters ($i,j,k, \ldots$) denote spatial indices and Greek indices ($\alpha, \beta, \ldots$) label different fluid species.
We consider flat
spatial slices in agreement with current observations \cite{Ade:2013zuv} with the matter content of the universe to be well-modelled by a perfect fluid, for which the energy-momentum tensor takes the form
\be 
T^\mu{}_\nu=(\rho+P)u^\mu u_\nu+P \delta^\mu{}_\nu\,.
\ee
Here, $\rho$ and $P$ are the energy density and pressure of the fluid, respectively, and $u^\mu$ is the fluid four-velocity, subject to the constraint $u_\mu u^\mu =-1$.

All perturbed quantities are then expanded in a series up to second order (following, e.g., Refs.~\cite{MW2008, thesis}) as, for example for the energy density,
\be 
\drho(x^i,\eta)=\delta\rho_1(x^i, \eta)
+\frac{1}{2}\delta\rho_2(x^i, \eta)+\cdots\,,
\ee
where the subscript denotes the order of the perturbation. The components of the fluid four-velocity are, up to second order in perturbation theory, then
\bea
&u_0=-a\Big[1+\phi_1+\frac{1}{2}\phi_2-\frac{1}{2}\phi_1^2+v_{1k}v_1^k\Big]\,,\\
&u_i=a\Big[V_{1i}+\frac{1}{2}V_{2i}-\phi_1 B_{1i}\Big]\,,\\
&u^0=\frac{1}{a}\Big[1-\phi_1-\frac{1}{2}\phi_2+\frac{3}{2}\phi_1^2+v_{1k}(B_1^k+V_1^k)\Big]\,,\\
&u^i=\frac{1}{a}\Big[v_1^i+\frac{1}{2}v_2^i\Big]\,,
\eea
where $v^i$ is the fluid three-velocity and $V^i=v^i+B^i$.

The governing equations are then the energy-momentum conservation and Einstein equations, respectively,
\bea
&\nabla_\mu T^\mu{}_\nu=0\,,\\
&G^{\mu}{}_\nu=8\pi G T^\mu{}_\nu\,.
\eea%
To solve these equations we perturb them to the required order, 
for this work up to second order. We do not present the equations in
detail here, but note that they can be found in, e.g.,
Ref.~\cite{thesis}.

\subsection{Magnetic fields and Maxwell equations}

The electromagnetic fields are described invariantly by the antisymmetric Faraday tensor, $F_{\mu \nu}$. We can then define fields as measured by a comoving observer: the electric field is
\be 
\label{eq:defE}
\Et^\mu = F^{\mu\nu} u_\nu\,,
\ee
and the magnetic field is\footnote{We choose to denote the magnetic field as $\Mt^\mu$ to avoid confusion with the metric perturbation $B^i$ which is non-zero in the uniform curvature gauge in which we work.}
\be 
\label{eq:defM}
\Mt^\mu = 
\frac{1}{2} \epsilon^{\mu\nu\lambda\delta} u_\nu F_{\mu\delta}\,,
\ee
where $\epsilon^{\mu\nu\lambda\delta}$ is the fully antisymmetric tensor, and
\bea
&\Et_\mu u^\mu=0\,,\\
&\Mt_\mu u^\mu =0\,.
\eea
 The Maxwell equations govern the evolution of the electromagnetic field and are written, in a compact form, as (e.g., \cite{Tsagas:2004kv})
\bea
&F_{[\mu\nu;\lambda]}=0\,,\\
&F^{\mu\nu}{}_{;\nu}=\mu_0j^\mu\,,
\eea
where $j^\mu=\frac{1}{a}(\hat{\rho},\mathbf{j})$ is the four-current that sources the electromagnetic field, $\hat{\rho}$ is the comoving charge density, $\mathbf{j}$ is the comoving three-current and $\mu_0$ is the magnetic permeability of the vacuum.

In order to perform the decomposition of the Maxwell equations, we introduce the projection tensor $h_{\mu\nu}$ defined as
\be 
h_{\mu\nu}=g_{\mu\nu}+u_\mu u_\nu\,,
\ee
which satisfies the conditions $h_{\mu\nu}u^\nu=0$, $h^\mu{}_\mu=3$ and $h^\mu{}_\nu h^\nu{}_\lambda=h^\mu{}_\lambda$. With this, the derivative of the fluid four-velocity can be decomposed as
\be
\nabla_\nu u_{\mu} = \sigma_{\mu\nu} + \omega_{\mu\nu} + \frac{1}{3}\theta h_{\mu\nu} - \dot{u}_{\mu}u_{\nu} \label{decu}
\ee
where $\dot{u}_{\mu} = u_{\mu;\nu}u^{\nu}$ and $\dot{u}_{\mu}u_{\mu} = 0$, $\omega_{\mu\nu}$ is an antisymmetric tensor and 
$\sigma_{\mu\nu} + \frac{1}{3}\theta h_{\mu\nu}$ is a symmetric tensor, with $\sigma_{\mu\nu}$ trace free and $\theta$ is the expansion scalar, $\theta=\nabla^\mu u_\mu$.
The four current can then be decomposed as 
\be
{\hat{\rho}} = -j^{\mu}u_{\mu}, \quad  \J^{\mu} = {h^{\mu}}_{\nu}j^{\nu} \,.  
\ee

We can now decompose the Maxwell equations by projecting along and orthogonal to the fluid four-velocity, $u^\mu$. In order to achieve this, we multiply the Maxwell equations by $u_\mu$ and $h^\mu{}_\nu$, respectively. We omit the working, and instead quote the result. We obtain two constraint equations, 
\bea
\label{Maxwell1}
&{\Et^{\mu}}_{,\mu} + \Gamma^{\mu}_{\kappa\mu} \Et^\kappa - \dot{u}_{\mu}\Et^{\mu} = \hat{\rho} - 2\omega^{\mu} \Mt_{\mu} \,,\\
&{\Mt^{\mu}}_{,\mu} + \Gamma^{\mu}_{\kappa\mu} \Mt^\kappa - \dot{u}_{\mu}\Mt^{\mu} = - \omega^{\mu} \Et_{\mu}\,,
\eea
and two evolution equations
\bea
&h^{\lambda}_{\mu}u^{\alpha}{\Et}^{\mu}_{,\alpha} = -(u^{\lambda} u_{\mu} {\Gamma}^{\mu}_{\kappa \alpha} - {\Gamma}^{\lambda}_{\kappa \alpha}) u^{\alpha} \Et^{\kappa}
+ (\omega^{\lambda}_{\nu} + 
\sigma^{\lambda}_{\nu} - \frac{2}{3} \theta h^{\lambda}_{\nu})\Et^{\nu} \nonumber\\
&\qquad+ \epsilon^{\lambda\nu\mu}\dot{u}_{\nu}\Mt_{\mu}
- \epsilon^{\lambda\nu\mu}(\Mt_{\nu,\mu}-\Gamma^\kappa_{\nu\mu}\Mt_\kappa) - \J^{\lambda} \,,  \\
&h^{\lambda}_{\mu}u^{\alpha}{\Mt}^{\mu}_{,\alpha} = -(u^{\lambda} u_{\mu} {\Gamma}^{\mu}_{\kappa \alpha} - {\Gamma}^{\lambda}_{\kappa \alpha}) u^{\alpha} \Mt^{\kappa} 
+ (\omega^{\lambda}_{\nu} + 
\sigma^{\lambda}_{\nu} - \frac{2}{3} \theta h^{\lambda}_{\nu})\Mt^{\nu} \nonumber\\
&\qquad\qquad- \epsilon^{\lambda\nu\mu}\dot{u}_{\nu}\Et_{\mu}
+ \epsilon^{\lambda\nu\mu}(\Et_{\nu,\mu}-\Gamma^\kappa_{\nu\mu}\Et_\kappa) 
\eea
where $\epsilon^{\lambda\nu\mu}=\epsilon^{\lambda\nu\mu\delta}u_\delta$ and we have used the fact that the covariant derivative of a vector is given by ${\Et^{\mu}}_{;\nu} = {\Et^{\mu}}_{,\nu} + \Gamma^{\mu}_{\kappa\nu} \Et^\kappa$.  Here, $\Gamma^\mu_{\nu\gamma}$ are the Christoffel symbols for perturbed FLRW and an overdot denotes a covariant derivative along the fluid flow, i.e. $\dot{u}_{\mu}=\nabla_\nu u_{\mu}u^\nu$. The vorticity vector is defined as
\be 
\label{eq:vortdef}
\omega^\mu=\epsilon^{\mu\nu\lambda}\omega_{\nu\lambda}\,.
\ee

\subsection{Maxwell equations in perturbation theory}

Having introduced cosmological perturbation theory along with the Maxwell equations in a covariant form, we are now in a position to combine the two, and to present the governing equations for an electromagnetic field in cosmological perturbation theory. Since neither the magnetic field \cite{Gopal:2004ut} nor the vorticity \cite{vorticity, Christopherson:2010dw} is not sourced in linear perturbation theory, we set $\Mt_{1i}$ and $\omega_1^i$ to zero, along with the linear shear.

Expanding the equations in the previous section up to linear perturbations results in the evolution and constraint equations for the electric field:
\bea
\label{eq:EIev}
&{{\EI}^{i}}'+ 2\H {\EI}^{i}  = -a{\mu_0 }{\J_1}^{i}\,,\\
\label{eq:EIcons}
& {\partial}_{i}{{\EI}^{i}} = {\mu_0}\hat{\rho_1}\,,
\eea
where a prime denotes a derivative with respect to conformal time, $\eta$.

To second order in perturbation theory, we obtain a set of equations for the electric field
\bea
& {\partial}_{i}{\EII}^{i}  +2\Big[2{\phi_{1,i}} - {{v_1}_{i}}'  - {{V_1}_{i}}' + 2\H{v_1}_{i} \Big] {\EI}^{i}
\label{eq:EIIcons}
 = \mu_0 \rho_2  - 2 a  \mu_0 {v_1}_{i}{\J_1}^{i}\,,\\
& {{\EII}^{i}}'  +2  \H  {\EII}^{i} = - 2a{\mu_0 }\phi_1{\J_1}^{i} - 2{v_1}^{j} {\partial}_{j}{{\EI}^{i}} \nonumber\\
\label{eq:EIIev}
&\qquad\qquad\qquad + \frac{4}{3} {\partial}_{j}{{v_1}^{j}} {\EI}^{i}  +{\epsilon}^{0 i j k}a^2{\partial}_{j}{{\MII}_{k}}  - a{ \mu_0 } {\J_2}^{i}\,,
\eea
along with the following pair of equations for the magnetic field
\bea
\label{eq:MIIcons}
& {\partial}_{i}{\MII}^{i} = 0\,,\\
\label{eq:MIIev}
&  {{\MII}^{i}}'  +2 \H {\MII}^{i}   = {\epsilon}^{0 i j k} a^2
\Big[2 \Big({\partial}_{j}{\phi_1} -{{V_1}_{j}}'+ 2{V_1}_{j} \H\Big) {\EI}_{k} \nonumber\\
&\qquad\qquad\qquad\qquad - {\partial}_{j}{{\EII}_{k}} 
 + 2 \mu_0 {V_1}_{j} a {\J_1}_{k}\Big]\,.
\eea

In order to close the system, we require equations governing the matter and gravity sector. These come from the Einstein field equations and energy-momentum conservation equations, as described above. In particular, the linear momentum conservation for a fluid, $\alpha$, is \cite{ks, Malik:2002jb}, where from now on we neglect linear vector perturbations so that ${V_1}_i=\partial_i V_1$,
\be
{{V_1}_{\alpha}}' +  (1 - 3 c_{\alpha}^2) \H {V_1}_{\alpha} + \phi_1 + \frac{1}{{\rho_0}_{\alpha} + {P_0}_{\alpha}}
\Big[{\dP_1}_{\alpha}  - \displaystyle\sum\limits_{\beta} {f_1}_{\alpha \beta }\Big] = 0\,,
\ee
where $c_\alpha^2$ is the adiabatic sound speed of the $\alpha$ fluid, i.e. $c_\alpha^2={P_0}_\alpha ' / {\rho_0}_\alpha '$ and $f_{\alpha\beta}$ is the momentum transfer between fluids \cite{Malik:2002jb}.

We consider a system containing three fluid species: protons (p), electrons (e) and photons ($\gamma$), with an electromagnetic background (F). The protons and electrons are assumed to act as pressureless matter,
hence $P_{\rm e} = P_{\rm p} = c_{\rm p} = c_{\rm e} = 0$, and the photons act as radiation so that $c_{\gamma}^2 = 1/3$. The linearly perturbed Einstein equations give us a constraint between the metric potential $\phi_1$ and the fluid velocities,
\be 
\phi_1 = - \frac{4\pi G a^2}{\H}\Big({\rho_0}_{\rm p} {V_1}_{\rm p} + {\rho_0}_{\rm e} {V_1}_{\rm e} + \frac{4}{3} {\rho_0}_{\gamma} {V_1}_{\gamma}\Big)\,.
\ee

Putting these together results in the following system of equations for the velocities of the fluid species
\bea
\label{eq:Vp}
&\fl
{V_1}_{\rm p}' + \H {V_1}_{\rm p} - \frac{3 \H}{2 \rho_0}\Big({\rho_0}_{\rm p} {V_1}_{\rm p} + {\rho_0}_{\rm e} {V_1}_{\rm e} + \frac{4}{3} {\rho_0}_{\gamma} {V_1}_{\gamma}\Big) 
- \frac{a}{{\rho_0}_{\rm p}} ({f_1}_{\rm pe} + {f_1}_{{\rm p}\gamma} +{f_1}_{\rm p F}) 
=  0\,, \\
\label{eq:Ve}
&\fl {V_1}_{\rm e}' + \H {V_1}_{\rm e} - \frac{3 \H}{2 \rho_0}\Big({\rho_0}_{\rm p} {V_1}_{\rm p} + {\rho_0}_{\rm e} {V_1}_{\rm e} + \frac{4}{3} {\rho_0}_{\gamma} {V_1}_{\gamma}\Big)
- \frac{a}{{\rho_0}_{\rm e}} ({f_1}_{\rm ep} + {f_1}_{{\rm e}\gamma}+{f_1}_{\rm e F} ) 
= 0\,, \\
\label{eq:Vgamma}
&\fl
{{V_1}_{\gamma}}'- \frac{3 \H}{2 \rho_0}\Big({\rho_0}_{\rm p} {V_1}_{\rm p} + {\rho_0}_{\rm e} {V_1}_{\rm e}+ \frac{4}{3} {\rho_0}_{\gamma} {V_1}_{\gamma}\Big) +\frac{1}{4\rho_{0\gamma}}\delta\rho_{1\gamma}-\frac{3a}{4\rho_{0\gamma}}(f_{1\gamma \rm p}+f_{1\gamma \rm e})=0\,.
\eea
The interaction terms between the species depend on the velocity difference, i.e. ${f_1}_{\alpha\beta}=\alpha_{\alpha\beta}({V_1}_{\alpha}-{V_1}_{\beta})$, where $\alpha_{\alpha\beta}$ are the interaction coefficients between the fluid species, and the momentum transfer with the electromagnetic field, to first order, is ${f_1}_{\rm sF}=q_{\rm s}n_{\rm s}\EI$. Substituting for these,
using the values for the constants found in the appendix, closes the system of equations.

\section{Results}
\label{sec:results}

Having introduced the formalism and presented our set of equations in the previous section, we are now in a position to solve the system. 
In order to achieve our goal to compute the second order magnetic field power spectrum, we must solve Eq.~(\ref{eq:MIIev}).
Assuming no vector perturbations and working, now, at an early time in a radiation background (where $10^{-12}<a<10^{-5}$), we can simplify the evolution equation for the second order magnetic field, Eq.~(\ref{eq:MIIev}), 
by using the governing equations, to become
\bea
\fl
 {{\MII}^{i}}'  + 2\H {\MII}^{i}   = 2a^2{\epsilon}^{0 i j k} 
\Bigg[\Big(\frac{\delta P_{1,j}}{c^2 \rho_0(1+w)} -(1-6c_{\rm s}^2+3w)\frac{V_{1,j}}{c}\Big){\EI}_{,k}
-ac\mu_0{\J_1}_{,k}V_{1,j}-\frac{1}{2}{{\EII}_{k,j}}\Bigg]\,,
\eea
which we denote, in a shorthand, as 
\be 
{{\MII}^{i}}'  +2 \H {\MII}^{i}   = S^i\,,
\ee
where $S^i$ is the source term for the equation. Here, we have introduced the equation of state parameter, $w=P_0/ \rho_0$ and the adiabatic sound speed $c_{\rm s}^2=P_0' / \rho_0'$. We can then transform to Fourier space, and on substituting for $V_1$ and dropping the term involving $\EII$, since it can be shown not to contribute to the source term, obtain
\bea
\fl
S^i (\mathbf{k} , \eta) =  \frac{a^2 }{(1+w)\rho_0}  
\frac{{\epsilon}^{0 i j k}k_k}{(2\pi)^{3/2}}
 \int d^3 \tilde{\mathbf{k}} \frac{\tilde{k}_j}{(9\H^2(1+w)+2c^2\tilde{k}^2)}\nonumber\\
\fl
\quad \times\Bigg[2ac^2\mu_0\Big(2\delta\rho_1'(\tilde{\bf k},\eta)+2\H(3+w)\delta\rho_1(\tilde{\rm k},\eta)
 +\frac{6\H}{c^2}\delta P_1(\tilde{\bf k},\eta)\Big)\J_1({\bf k}-{\bf \tilde{k}},\eta)\nonumber\\
\fl\qquad\qquad-\Bigg\{\H(1-6c_{\rm s}^2+3w)\Big(2\delta\rho_1'(\tilde{\bf k},\eta)+3\H(3+w)\delta\rho_1(\tilde{\bf k},\eta)\Bigg)\nonumber\\
\fl \qquad\qquad-\frac{4}{c^2}\Big(3\H^2(3c_{\rm s}^2+1)+c^2\tilde{k}^2\Big)\delta P_1(\tilde{\bf k},\eta)\Big\}\EI({\bf k}-\tilde{\bf k},\eta)\Bigg]
 \eea

In order to solve this, we follow the calculation in Refs.~\cite{Christopherson:2010ek, thesis}, and expand the magnetic field vector by employing the basis
\be 
{\Mt}_i({\bm k}, \eta)={\Mt}_A({\bm k}, \eta)e_i({\bm k})+{\Mt}_B({\bm k},\eta)\bar{e}_i({\bm k})+{\Mt}_C({\bm k},\eta)\hat{k_i}\,,
\ee
where the subscripts $A, B, C$ denote the three Fourier modes. Noting that the magnetic field, like the vorticity, is an axial vector, we find that
\bea
\fl
 S_A (\mathbf{k} , \eta) = - \frac{a^2 }{(1+w)\rho_0}  
\frac{k \bar{e}^j}{(2\pi)^{3/2}}
 \int d^3 \tilde{\mathbf{k}} \frac{\tilde{k}_j}{(9\H^2(1+w)+2c^2\tilde{k}^2)}\nonumber\\
\fl
\quad \times\Bigg[2ac^2\mu_0\Big(2\delta\rho_1'(\tilde{\bf k},\eta)+2\H(3+w)\delta\rho_1(\tilde{\rm k},\eta)
 +\frac{6\H}{c^2}\delta P_1(\tilde{\bf k},\eta)\Big)\J_1({\bf k}-{\bf \tilde{k}},\eta)\nonumber\\
\fl\qquad\qquad -\Bigg\{\H(1-6c_{\rm s}^2+3w)\Big(2\delta\rho_1'(\tilde{\bf k},\eta)+3\H(3+w)\delta\rho_1(\tilde{\bf k},\eta)\Bigg)\\\nonumber
\fl \qquad\qquad-\frac{4}{c^2}\Big(3\H^2(3c_{\rm s}^2+1)+c^2\tilde{k}^2\Big)\delta P_1(\tilde{\bf k},\eta)\Big\}\EI({\bf k}-\tilde{\bf k},\eta)\Bigg]\,,\\
\fl
 S_B (\mathbf{k} , \eta) = \frac{a^2 }{(1+w)\rho_0}  
\frac{ke^j}{(2\pi)^{3/2}}
 \int d^3 \tilde{\mathbf{k}} \frac{\tilde{k}_j}{(9\H^2(1+w)+2c^2\tilde{k}^2)}\nonumber\\
\fl
\quad \times\Bigg[2ac^2\mu_0\Big(2\delta\rho_1'(\tilde{\bf k},\eta)+2\H(3+w)\delta\rho_1(\tilde{\rm k},\eta)
 +\frac{6\H}{c^2}\delta P_1(\tilde{\bf k},\eta)\Big)\J_1({\bf k}-{\bf \tilde{k}},\eta)\nonumber\\
\fl \qquad\qquad-\Bigg\{\H(1-6c_{\rm s}^2+3w)\Big(2\delta\rho_1'(\tilde{\bf k},\eta)+3\H(3+w)\delta\rho_1(\tilde{\bf k},\eta)\Bigg)\nonumber\\
\fl\qquad\qquad -\frac{4}{c^2}\Big(3\H^2(3c_{\rm s}^2+1)+c^2\tilde{k}^2\Big)\delta P_1(\tilde{\bf k},\eta)\Big\}\EI({\bf k}-\tilde{\bf k},\eta)\Bigg]\,,\\
\fl
  S_C (\mathbf{k}, \eta)  =  0 \,.
\eea
The two point correlator of the magnetic field is then computed from the source term as 
 \be
 \fl
 \langle \M^{*}(\mathbf{k_1},\eta) \M(\mathbf{k_2},\eta) \rangle = 
 \eta^{-4} \int_{\eta_0}^{\eta} d \tilde{\eta_1} \tilde{\eta_1}^2 \int_{\eta_0}^{\eta} d \tilde{\eta_2} \tilde{\eta_2}^2
 \langle S^{*} (\mathbf{k_1} ,\eta_1) S(\mathbf{k_2} ,\eta_2) \rangle\,.
 \ee
 
 We now focus on the $S_A$ term, since the amplitudes of the two non-zero polarisations are identical, up to the basis vector (dropping the subscript in the following), and work on large scales, using the approximation $c^2k^2 << 6\H^2$. Furthermore, assuming that the electric field and current can be decomposed into an $\eta$-dependent and $k$-dependent piece, we note from Eq.~(\ref{eq:EIev}) that the scale dependence of $\EI$ and $\J_1$ are identical, and therefore
 \bea
 \J_1({\bf k}, \eta)=J(\eta)\EI({\bf k})\,,\\
 \EI({\bf k}, \eta) = E(\eta)\EI({\bf k})\,.
 \eea
 Thus, the source term can be written in a simplified form as
 \bea
 \label{eq:Ssimp}
 \fl
 S({\bf k},\eta)=\frac{a^2 k \bar{e}^j}{(1+w)\rho_0}
 \int\frac{d^3{\bf k} \tilde{k}_j}{9\H^2(1+w)+2c^2\tilde{k}^2}
 \Big[f(\tilde{k},\eta)\delta\rho_1(\tilde{\bf k},\eta)
 +g(\tilde{k},\eta)\delta P_{\rm nad 1}(\tilde{\bf k},\eta)\Big]
 \EI({\bf k}-\tilde{\bf k})\,,
 \eea
 where we have introduced the functions
 \bea
 f(\tilde{k},\eta)&\equiv 2\H ac^2\mu_0(1+3w+6 c{\rm s}^2)J(\eta)\\
 &\qquad-\Big(\H^2\Big[(1-6 c_{\rm s}^2+3w)(1+3w)-3c_{\rm s}^2(3c_{\rm s}^2+1)\Big]-c^2c_{\rm s}^2\tilde{k}^2\Big) E(\eta)\,,\nonumber\\
 g(\tilde{k},\eta)&\equiv \frac{4}{c^2}\Big(3\H ac^2\mu_0 J(\eta)
 +\Big[3\H^2(3c_{\rm s}^2+1)+c^2\tilde{k}^2\Big]E(\eta)\,,
 \eea
 and have split the pressure perturbation as 
 \be 
 \label{eq:defdPnad}
 \delta P_1 = c_{\rm s}^2\delta\rho_1+\delta P_{\rm nad 1}\,,
 \ee
 where $\delta P_{\rm nad 1}$ is the non-adiabatic pressure perturbation.
 
 In order to complete this calculation, we now need to obtain solutions for the energy density and pressure perturbations and the electric and magnetic field, via the velocity differences. This will be the focus of the next subsections.
 
 \subsection{ Energy density and pressure perturbations}
\label{sec:drho}

The solutions for the linear energy density and pressure are well known. At early times and on large scales, the solution for the density perturbation is \cite{Christopherson:2010ek} (where we have dropped the subscript for this section, since we are considering linear energy density and pressure perturbations)
\be 
\delta\rho_{\gamma}({\bf k},\eta)={A({\bf k})}\Big(\frac{\eta}{\eta_0}\Big)^{-4}\,.
\ee
The scale dependence can then be determined from observations. 
We know that
\be
A = \drho_{\rm{init}} \left(\frac{k}{k_0}\right)^{\frac{1}{2}(1-n_{\rm s})} \,,
\ee
and the energy density perturbation in the flat gauge can be related to the curvature perturbation on uniform density hypersurfaces, $\zeta$, during radiation domination through 
\be
\drho = -\frac{\rho_0'\zeta}{\H} = 4\rho_0 \zeta\,,
\ee
and hence the initial power spectra can be related as
$
\langle \drho_{\rm{init}} \drho_{\rm{init}} \rangle = 16 \rho^2_{0 \rm{init}} \langle \zeta_{\rm{init}} \zeta_{\rm{init}} \rangle\,,
$
where
\be
\fl
\langle \zeta_{\rm{init}} \zeta_{\rm{init}} \rangle = \frac{2\pi}{k^3} \Ps_{\zeta} (k, \eta_{\rm{init}}) = \frac{2\pi^2}{k^3} L^3 \Delta^2_{\zeta}(k) = \frac{2\pi^2}{k^3} L^3 \Delta^2_{\zeta}(k_0) 
 \left(\frac{k}{k_0}\right)^{n_s -1} \,,
\ee
where we have introduced the length scale $L$  to correct the units. 
Substituting this into the above we have
\be
A^2 = 32 \pi^2 \rho_{0\rm{init}}^2 L^3 k_0^{-3} \Delta^2_{\R}(k_0)   \,,
\ee
which will prove to be a required amplitude later.

In order to solve for the pressure, we use the non-adiabatic pressure perturbation defined above in Eq.~(\ref{eq:defdPnad}).
Since we know the behaviour of the density perturbation, we focus on the non-adiabatic part of the pressure perturbation. Each individual fluid is assumed to be a perfect fluid, and so does not have an intrinsic non-adiabatic part. However, there are two other origins of non-adiabatic pressure in our system. These are: (\emph{i}) the non-adiabatic pressure perturbation which arises from inflation drive by multiple fields and imprinted as an isocurvature fraction in the CMB ($\delta P_{\rm inf}$), and (\emph{ii})
the relative non-adiabatic pressure perturbation caused by the interaction between the different fluids ($\delta P_{\rm rel}$). 

The inflationary contribution is close to scale-invariant, and has the functional form
\be 
\delta P_{\rm inf}=D_{\rm inf}\Big(\frac{\eta}{\eta_0}\Big)\,,
\ee
while the relative contribution has the approximate solution at early times and on large scales \cite{Brown:2011dn}
\be
\delta P_{\rm rel}= D_{\rm rel} \Big(\frac{k}{k_0}\Big)^4 \Big(\frac{\eta}{\eta_0}\Big)\,.
\ee
Since these are both power law scalings, we will use the following expression throughout our calculation in order to accommodate both cases,
\be
\delta P_{\rm nad}=P\Big(\frac{k}{k_0}\Big)^m \Big(\frac{\eta}{\eta_0}\Big)\,,
\ee
where $P$ and $m$ depend on which of the above cases we are interested in.

In order to obtain $D_{\rm inf}$, we consider the non-adiabatic pressure perturbation. The comoving entropy perturbation introduced in Refs.~\cite{Gordon:2000hv, seminal} is defined as
\be 
{\mathcal S}=\frac{{\mathcal H}}{c^2P'}\delta P_{{\rm inf}}\,,
\ee
which, in a radiation background, reduces to 
\be 
\delta P_{{\rm inf}}=-\frac{4}{3}c^2\rho_0{\mathcal S}\,.
\ee
From the definition of the entropy power spectrum, we can relate the power in the curvature perturbation to the power in the isocurvature perturbation through the function $\alpha(k)$,
\be
\frac{\alpha(k_0)}{1 - \alpha(k_0)} = \frac{\Ps_S(k_0)}{\Ps_{\R}(k_0)}
\equiv\hat{\alpha}^2\,,
\ee
where we note the standard definitions for the power spectrum
\bea
\Ps_{\R}(k, \eta) &= \frac{k^3}{2\pi} \langle | \R(k, \eta) |^2 \rangle\nn\\
 \Delta_{\R}^2 (k) &= \frac{k^3}{2 \pi^2 L^3} \langle | \R(k) |^2 \rangle = \Delta_{\R}^2 (k_0) \left( \frac{k}{k_0} \right)^{n_s - 1}\,.
 \eea
 We can then write the entropy power spectrum as
 \be
 \Ps_{S}(k, \eta) = \frac{k^3}{2\pi} \langle | S(k, \eta) |^2 \rangle 
 = \frac{\alpha(k_0)}{1 - \alpha(k_0)} \pi L^3 \Delta_{\R}^2 (k_0) \left( \frac{k}{k_0} \right)^{n_s - 1} \,.
 \ee
Combining these, we obtain
\be 
D_{\rm inf}^2=\rho_{0 \rm init}^2 c^4 \frac{32\pi^2}{9k_0^3}L^3\frac{\alpha(k_0)}{1-\alpha(k_0)}\Delta^2_{\mathcal R}(k_0)\,.
\ee

The amplitude for the relative non-adiabatic pressure perturbation, $D_{\rm rel}$, is obtained from Ref.~\cite{Brown:2011dn} as approximately $10^{-3} {\rm Mpc}^{-1}$.

\subsection{Velocity differences, current and electric field}

We are interested in obtaining a solution for the magnetic field around recombination, where the tight coupling approximation breaks down. This means that the protons and electrons move independently, and so $V_{1\rm e}\neq V_{1\rm p}$. Additionally, since recombination occurs after matter-radiation equality, we cannot assume a background of radiation when computing the velocity differences; instead, we introduce the baryon to photon ratio, $R_{\rm b}$.

In order to solve the above set of equations for the velocity difference, we assume that the time dependence of the three velocities is well-described as a power law, e.g., $V_{1\gamma}=\hat{V}_{1\gamma}(x^i)\eta^n$. Then, the set of Eqs.~(\ref{eq:Ve}), (\ref{eq:Vp}), (\ref{eq:Vgamma}), together with the definition for the linear current in terms of the velocity difference of protons and electrons,
\be
{\J_1} =  c a e n (V_{1 \rm p} - V_{1 \rm e})\,,
\ee
can be solved, employing the approximation for the energy density perturbation of radiation, presented in Section~\ref{sec:drho}.\footnote{Although we do not want to assume radiation domination, we are only interested in the time up to and including recombination, and therefore we will still restrict the calculation to $a\leq 10^{-3}$. During this period, the factor $(1+4/3 \hat{R}_{b})$, which enters the calculation through the expression for the Hubble parameter, takes the range of values
\be 
1 < (1+4/3 \hat{R}_{b}) < 2\,.
\ee
Since we are interested in only an order of magnitude result for the final solution, we can safely approximate this to 1, which allows us to solve the system of equations.} Furthermore, we assume that the electric field and current have the same scale dependence, which is well-described as a power law, e.g., $\EI({\bf k}, \eta)=\bar{E}(\eta)k^l$ where, for the large scales on which we are working, $l=0$.

The solution for the velocity difference results in the following expression for the electric field
\be
\bar{E}(\eta)=\Bigg(\frac{2A\beta c^2}{3e\hat{n}_b}a^{-2}
+\frac{2A\sigma_{Te}^2c^2m_p\beta^2}{3\mu_0\hat{R}_be^3}a^{-8}\Bigg)\,,
\ee
where we have included the two most dominant terms. Using Eq.~(\ref{eq:EIev}), we can then obtain the linear current
\be 
\bar{J}(\eta)=\frac{4A\sigma_{Te}^2m_p\beta^2}{\mu_0^2\hat{R}_be^3\eta_0}a^{-10}\,.
\ee

\subsection{Power spectrum of the second order magnetic field}

We are now in a position to compute the power spectrum of the second order magnetic field, putting together the previous elements of the calculation. Recall that we are working on large scales, and in a radiation background. In this case, and noting that $ S^* (\mathbf{k}, \eta) =  - S (-\mathbf{k}, \eta)$, the source term Eq.~(\ref{eq:Ssimp}) then gives rise to the correlator
\bea
\label{eq:Scorrelator}
\fl
\langle S({\bf k}_1,\eta_1)S^*({\bf k}_2,\eta_2)\rangle
=\frac{\eta_1^8\eta_2^8}{(2\pi)^34^4\rho_0^2\eta_0^{12}}k_1 k_2\bar{e}^j\bar{e}^k
\delta({\bf k}_1- {\bf k}_2)\\
\fl
\qquad\times \int d^3\tilde{\bf k}\tilde{k}_k\Big[\tilde{k}_j\Big\{
f(\eta_1)f(\eta_2)A^2\eta_0^8\eta_1^{-4}\eta_2^{-4}
+f(\eta_1)g(\eta_2)AP\eta_0^3k_0^{-m}\eta_1^{-4}\eta_2\tilde{k}^m
\nonumber\\
\fl
\qquad\qquad\qquad
+g(\eta_1)f(\eta_2)AP\eta_0^3k_0^{-m}\eta_1\eta_2^{-4}\tilde{k}^m
+g(\eta_1)g(\eta_2)P^2\eta_0^{-2}k_0^{-2m}\eta_1\eta_2\tilde{k}^{2m}\Big\}
\nonumber\\
+|\tilde{\bf k}-{\bf k}|_j\Big\{f(\eta_1)f(\eta_2)A^2\eta_0^8\eta_1^{-4}\eta_2^{-4}
+f(\eta_1)g(\eta_2)AP\eta_0^3k_0^{-m}\eta_1^{-4}\eta_2
|\tilde{\bf k}-{\bf k}|^m\nonumber\\
+g(\eta_1)f(\eta_2)AP\eta_0^3k_0^{-m}\eta_1\eta_2^{-4}\tilde{k}^m
+g(\eta_1)g(\eta_2)P^2\eta_0^{-2}k_0^{-2m}\eta_1\eta_2\tilde{k}^m
|\tilde{\bf k}-{\bf k}|^m\Big\}\Big]\nonumber\,,
\eea
where we have used Wick's theorem and integrated out the delta functions, following the calculation in \cite{Christopherson:2010ek}
and the functions $f(\eta)$ and $g(\eta)$ are
\bea
f(\eta)=\frac{4Ac^2\beta}{3\eta_0^2\hat{n}_be}\Big(\frac{24\hat{n}_b\sigma_{Te}^2m_p\beta}
{\mu_0\hat{R}_be^2}-\eta^6\Big)\eta^{-10}
=E(J\eta_0^8\eta^{-10}-\eta_0^2\eta^{-4})\,,\\
g(\eta)=\frac{48A\beta}{3\eta_0^2\hat{n}_be}\Big(\frac{3\hat{n}_b\sigma_{Te}^2m_p\beta}
{\mu_0\hat{R}_b e^2}-\eta^6\Big)\eta^{-10}
=\frac{12E}{c^2}\Big(\frac{J}{8}\eta_{0}^8\eta^{-10}-\eta_0^2\eta^{-4}\Big)\,,
\eea
where we have introduced the constants
\bea 
E=\frac{4Ac^2\beta}{3\hat{n}_b e}\,,\\
J=\frac{24\hat{n}_b\sigma_{Te}^2m_p\beta}{\mu_0\hat{R}_be^2}\,.
\eea

In order to solve the integral in Eq.~(\ref{eq:Scorrelator}), we switch to spherical coordinates $(k,\theta,\varphi)$, for which the integral becomes
\bea
\fl
\int^{k_c}_0  d\tilde{k} \int^{\pi}_0 \sin^3\theta d\theta   
 \Bigg(
 f(\eta_1)f(\eta_2) A^2 \eta_0^8 \eta_1^{-4} \eta_2^{-4} k^4  \Bigg[\Big(\frac{\tilde{k}}{k}\Big)^4 + \left( 1 + \Big(\frac{\tilde{k}}{k}\Big)^2 - 2\Big(\frac{\tilde{k}}{k}\Big)\cos\theta \right)^{1/2} \Big(\frac{\tilde{k}}{k}\Big)^3 \Bigg] \nonumber \\
\fl
\qquad
 + f(\eta_1)g(\eta_2) A P\eta_0^3 k_0^{-m}  \eta_1^{-4} \eta_2  k^{m+ 4} 
\Bigg[  \Big(\frac{\tilde{k}}{k}\Big)^{m + 4} + \left( 1 + \Big(\frac{\tilde{k}}{k}\Big)^2 - 2\Big(\frac{\tilde{k}}{k}\Big)\cos\theta \right)^{\frac{1}{2}(m+1)} \Big(\frac{\tilde{k}}{k}\Big)^3 \Bigg] \nonumber \\
 \fl\qquad+ g(\eta_1)f(\eta_2) A P \eta_0^3 k_0^{-m} \eta_1 \eta_2^{-4} k^{m + 4} 
\Bigg[ \Big(\frac{\tilde{k}}{k}\Big)^{m + 4} +  \Big(\frac{\tilde{k}}{k}\Big)^{m + 3} \left( 1 + \Big(\frac{\tilde{k}}{k}\Big)^2 - 2\Big(\frac{\tilde{k}}{k}\Big)\cos\theta \right)^{1/2} \Bigg] \nonumber \\
\fl \qquad
+ g(\eta_1) g(\eta_2) P^2 \eta_0^{-2} k_0^{-2m} \eta_1  \eta_2  k^{2m + 4} 
\Bigg[ \Big(\frac{\tilde{k}}{k}\Big)^{2\delta + 4}    + \Big(\frac{\tilde{k}}{k}\Big)^{\delta + 3}  \left( 1 + \Big(\frac{\tilde{k}}{k}\Big)^2 - 2\Big(\frac{\tilde{k}}{k}\Big)\cos\theta \right)^{\frac{1}{2}(m +1)}    \Bigg]\Bigg)\,,
\eea
where we have introduced a small-scale cut-off such that $k<k_c$.\footnote{This cut-off is required since, on sufficiently small scales, the cosmological calculation we focus on in this paper will be dominated by strongly nonlinear astrophysical effects and so perturbation theory will break down.}
This integral is most easily computed using a further change of variables,
\be
v = \frac{\tilde{k}}{k}\,,\qquad
u^2 =  \Bigg(1 + \Big(\frac{\tilde{k}}{k}\Big)^2 -2 \Big(\frac{\tilde{k}}{k}\Big)\cos\theta\Bigg) \,,
\ee
for which we can write the correlator in the form
\bea   
\fl \langle S(\mathbf{k}_1, a_1)  S^*(\mathbf{k}_2, a_2)  \rangle
 = \frac{\pi \eta_1^{8} \eta_2^{8}}{(2\pi)^34^5 \rho_0^2  \eta_0^{12}}   
 \Big[
 f(\eta_1)f(\eta_2) A^2 \eta_0^8 \eta_1^{-4} \eta_2^{-4} I_1(k)
 + f(\eta_1)g(\eta_2) A P\eta_0^3 \eta_1^{-4} \eta_2  I_2(k)  \nonumber \\
\fl\qquad\qquad\qquad + g(\eta_1)f(\eta_2) A P \eta_0^3  \eta_1 \eta_2^{-4} I_3(k) 
+ g(\eta_1) g(\eta_2) P^2 \eta_0^{-2} \eta_1  \eta_2  I_4(k) \Big]
\delta(\mathbf{k}_1 - \mathbf{k}_2) \,,
\eea
where the individual integrals are
\bea
\fl 
I_1(k) &=& k^7 \int^{k_c/k}_0 
\int^{|1 + v|}_{|1 - v |}( v + u  )u \left( 4 v^2 - (1 + v^2 - u^2)^2 \right) d u d v \,, \\
\fl
I_2(k) &=& k^{m +7} k_0^{-m} 
\int^{k_c/k}_0 \int^{| 1 + v | }_{| 1 - v |}(  v^{m + 1} + u^{m +1} )u \left( 4 v^2 - (1 + v^2 - u^2)^2 \right) d u d v \,, \\
\fl
I_3(k) &=& k^{m +7} k_0^{-m} 
\int^{k_c/k}_0 \int^{| 1 + v | }_{| 1 - v |}( v^{m + 1} +  v^{m} u)u \left( 4 v^2 - (1 + v^2 - u^2)^2 \right) d u d v \,, \\
\fl
I_4(k) &=& k^{2m +7} k_0^{-2m} 
\int^{k_c/k}_0 \int^{| 1 + v | }_{| 1 - v |}(  v^{2m + 1}    + v^{m}  u^{m +1}    )u \left( 4 v^2 - (1 + v^2 - u^2)^2 \right) d u d v \,. 
\eea
The solution of these integrals depends on which source of non-adiabatic pressure we are considering, as discussed above.
The time integrals can then be evaluated to give the following expression for the power spectrum of the magnetic field
\bea
\label{eq:ps}
\fl
k^3 \Ps_{\M}(k,\eta)=\frac{k^6}{2(2\pi)^3 4^5\rho_0^2}E^2\eta^{-4}\eta_0^{-6}\Bigg[\frac{A^2}{9}I_1(k)
+\frac{AP}{2c^2}\Big(I_2(k)+I_3(k)\Big)
+\frac{9P^2}{4c^4}I_4(k)\Bigg]\,.
\eea

Since we are interested in the magnitude of the magnetic field, we consider $\sqrt{k^3 \Ps_{\M}}$. Substituting the above expression for the amplitudes, in turn, into Eq.~(\ref{eq:ps}), along with numerical values for the constants (given in the appendix), keeping only the leading order term, and converting the units into Gauss, we obtain, first for the inflationary non-adiabatic pressure
 \bea
 \fl
 \sqrt{k^3 \Ps_{\M} (k, \eta)} &=& \frac{A E \eta_0}{32 \sqrt{2} (2\pi)^{3/2} \rho_0}  \left(\frac{k_c}{Mpc^{-1}}\right)^{\frac{13}{2}}  \left(\frac{\eta_c}{\eta}\right)^2 \left[ \frac{32 }{135}  + \hat{\alpha} \frac{16}{27}\frac{k_c}{k_0} + \hat{\alpha}^2 \frac{8}{21} \left(\frac{k_c}{k_0}\right)^2 \right]^{\frac{1}{2}} \left( \frac{k}{k_c} \right)^4  \,,
 \eea
and for the relative non-adiabatic pressure, the magnetic field power spectrum is
\bea
\fl
 \sqrt{k^3 \Ps_{\M} (k, \eta)} &=& \frac{E A \eta_0}{32 \sqrt{2}(2\pi)^{3/2}\rho_0}  \left(\frac{k_c}{Mpc^{-1}}\right)^{\frac{13}{2}} \left(\frac{\eta_c}{\eta}\right)^2  \left[\frac{32}{135} + \frac{32}{27}\frac{\hat{D}}{A}\left(\frac{k_c}{k_0}\right)^4 + \frac{24}{13}\frac{\hat{D}^2}{A^2} \left(\frac{k_c}{k_0}\right)^8\right]^{\frac{1}{2}}
\left(\frac{k}{k_c}\right)^4 \,,
\eea
where $\hat{D}=D_{\rm rel}/c^2$.

 As expected, this result depends on our small scale cut-off, $k_{c}$, and both sources of non-adiabatic pressure result in a field which scales like ${\mathcal{M}}\propto k^4\eta^{-2}$, in agreement with other work \cite{Fenu:2010kh}. We now take the cut-off scale to be $k_c = 10 Mpc^{-1}$ for illustrative purposes, and evaluate the spectrum from the inflationary contribution at $\eta = \eta_{\rm{eq}}$, this time including all terms from the $I(k)$ integrals above, instead of the dominant contributions, to obtain
\be
\fl
\sqrt{ k^3 \Ps_{\M} } =  3.2\times 10^{-17}\Bigg[736.3\left(\frac{k}{10}\right)^8+515.4\left(\frac{k}{10}\right)^{10}-\frac{4}{315}\left(\frac{k}{10}\right)^{12}+\frac{4}{2835}\left(\frac{k}{10}\right)^{14}\Bigg]^{\frac{1}{2}}\,.
\ee
We note that the power spectrum is rising towards smaller scales. 

Finally, we estimate the magnetic field strength, for both cases, on cluster scales of $k=1Mpc^{-1}$ and evaluated today. For the inflationary non-adiabatic pressure we obtain
\be
\sqrt{ k^3 \Ps_{\M} } \approx 5.9\times 10^{-27}G \,,
\ee
and for the relative non-adiabatic pressure
\be
\sqrt{ k^3 \Ps_{\M} } \approx 2\times 10^{-30}G \,.
\ee

Our results are heavily dependent on the cut-off scale, $k_c$, which is to be expected. We are limited in our choice of cut-off and although we would like to take the cut-off as high as possible (since the spectrum is rising) our series approximations are only valid in the regime $ak \ll 7630 Mpc^{-1}$. We also want the cut-off to be larger than the scales we are interested in, which are cluster scales ($k\sim 1Mpc^{-1}$). So, in quoting the result above, we choose $k_c=10Mpc^{-1}$, a reasonable value for both of these limits, in order to illustrate the results. 

\begin{figure}[ht!]
\centering
\includegraphics[width=120mm]{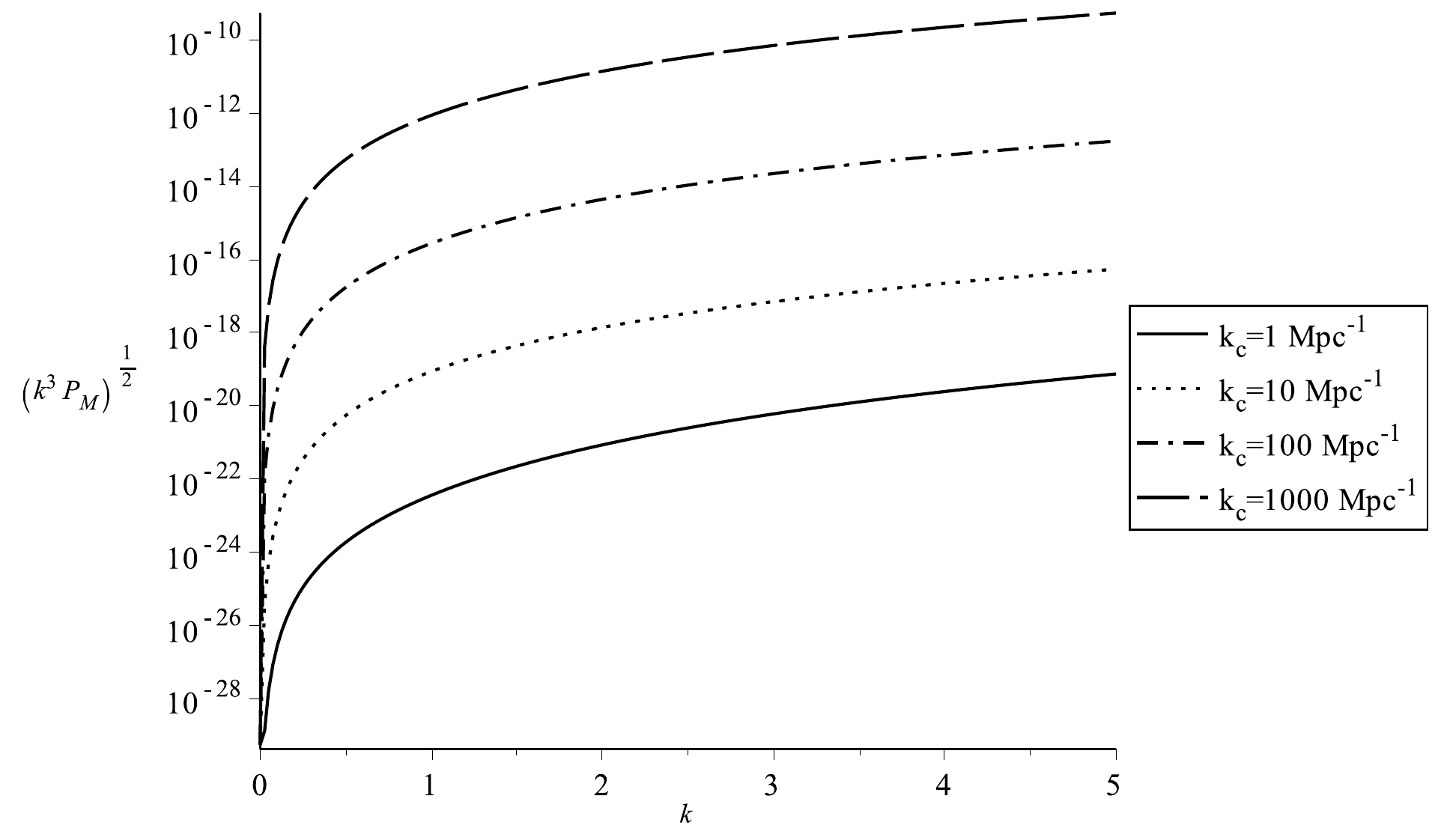} 
\caption{A plot showing $\sqrt{ k^3 \Ps_{\M} }$ in the scenario where we have inflationary non-adiabatic pressure for illustrative  choices of $k_c$ evaluated at $ \eta = \eta_{\rm{eq}}$.}
\label{fig:powerspec}
\end{figure}

If we vary the cut-off slightly between $k_c=1Mpc^{-1}$ to $k_c=1000Mpc^{-1}$ we get results that vary from $\sim 10^{-30} - 10^{-20} G$ (however we should not put too much trust in the upper end of the scale). The results for the inflationary non-adiabatic pressure (evaluated at matter-radiation equality) are plotted in Fig.~\ref{fig:powerspec}.

\section{Discussion}
\label{sec:dis}

In this paper we have revisited the topic of magnetic field generation
at second order in cosmological perturbation theory using solely
analytical techniques. This is a beneficial task, since it allows us
to understand the primordial magnetic field generated in the early
universe without having to rely on numerical computations.  We have
derived the equations governing the electromagnetic field using full
relativistic metric perturbation theory and presented the equations up
to second order.  By making simple approximations for the velocity
difference, we have then computed the current and the electric field.
Using expressions for the energy density and non-adiabatic pressure
perturbation from linear perturbation theory, we have then computed
the second order magnetic field on cluster scales, obtaining a
magnitude of $\sqrt{ k^3 \Ps_{\M} } \approx 5.9 \times 10^{-27} G$ and $\sqrt{ k^3 \Ps_{\M} } \approx 2 \times 10^{-30} G$, for our two cases, at
$k=1Mpc^{-1}$ with a scale dependence of $\sqrt{ k^3 \Ps_{\M} } \propto
k^4$, evaluated for the small scale cut-off value of $k_c=10Mpc^{-1}$.  The result depends on the small scale cut-off, and on choosing slightly different cut-off values, we obtain slightly different results, as quantified in the previous section. 

This is the first analytical calculation of the second order
magnetic field which takes into account all source terms in the
evolution equation.  Our result is in agreement with the relevant
numerical calculation presented in Ref.~\cite{Fenu:2010kh}. Since it
is well known that some Boltzmann codes have convergence issues, as
pointed out in Refs.~\cite{Huang:2012ub, Pettinari:2013he}, 
our analytical
calculation strengthens the numerical result and adds to the
literature on the magnetic field generated by second order effects. Additionally, the numerical calculations assume adiabatic initial conditions, and therefore do not taken into account any amplification due to the inflationary non-adiabatic pressure that we consider in our work.

Although the magnetic field we find from solely second order effects is perhaps too small to act as the primordial seed field, this should not be taken as the final 
word on the matter. 
As we have shown, the power spectrum is rising towards smaller scales in agreement with the result of the fully numerical calculation presented in Ref.~\cite{Fenu:2010kh}.
It is not impossible that power could move coherently from short to large scales and therefore a complete calculation including 
small scales could lead to an enhanced result for the amplitude of magnetic fields today. To see if this is indeed the case one would need to 
study the small scale result in more detail. This is beyond the scope of this paper and is left for future work.
We also recall that the origin of the first magnetic fields in our Universe is still largely unknown. Therefore, it is particularly important to continue to investigate 
the possibility that their origin is due to the non-linear nature of gravity, since this mechanism requires the introduction of no new physics. As described in the 
introduction, there are many different models of magnetogenesis in the very early universe that can generate a small seed field, each of which has its own problem. 
However, the calculations of the size of the magnetic field generated have all assumed that the field decays with the expansion of the universe (i.e. decays like radiation), 
after the magnetogenesis mechanism turns off. As presented in this article, on allowing for second order perturbations a magnetic field is generated. Therefore, in order 
to obtain a true prediction from these inflationary magnetogenesis mechanisms, non-linear effects must be included. For example, the magnetic field may not decay as quickly as the current estimates assuming a decay with radiation predict,
and the resultant field might be larger than predicted.
Using the analytical framework we have developed, we will investigate this interesting scenario in a future article \cite{magneto_nonlinear}.

Finally, it would be interesting to compare our results to additional numerical computations. There have been some great improvements in the sophistication of Boltzmann CMB codes to deal with perturbations beyond first order in the past year \cite{Huang:2012ub, Su:2012gt,  Pettinari:2013he}. Using these codes to perform a computation of the magnetic field both solely from non-linear effects and also including a non-zero linear order seed field will be an exciting task for the future. This will enable us to fully understand the magnetic field generated by non-linear cosmological perturbations.

\ack{
EN is funded by an STFC studentship, AJC by the Sir Norman Lockyer Fellowship of the Royal Astronomical Society and KAM is supported, in part, by STFC grant ST/J001546/1. {\sc Cadabra} \cite{Peeters:2007wn, Peeters:2006kp} was used in the derivation of some equations. AJC is grateful to the Astronomy Unit at QMUL for hospitality at various stages of the project. The authors are grateful for the helpful contributions and suggestions of the anonymous referee.}

\appendix

\section{Appendix}

\setcounter{section}{1}

\subsection{Interaction coefficients}

The interaction coefficients for the velocity difference equations are
\bea
\alpha_{\rm pe} &= \frac{n^2 e^2}{4\pi \epsilon_0 \sigma_C} =  \frac{n^2}{\sigma_C} (2.30707706 \times 10^{-28}) \rm{kg m^3 s^{-2}} \\
\alpha_{{\rm e} \gamma} &= \frac{4}{3} n c \sigma_T \rho_{\gamma} = \frac{n^2 c \sigma_T (m_{\rm p} +m_{\rm e})}{R_b} \nonumber\\
&=  \frac{n^2}{R_b} (3.33762112 \times 10^{-47}) \rm{kg m^3 s^{-1}}  \\
\alpha_{{\rm p} \gamma} &= \frac{4 \beta^2}{3 } n c \sigma_T \rho_{\gamma} = \frac{\beta^2 n^2 c \sigma_T (m_{\rm p} + m_{\rm e})}{R_b}\nonumber\\
&=  \frac{n^2}{R_b} (9.89964136 \times 10^{-54}) \rm{kg m^3 s^{-1}}\,.
\eea

Noting that $n_{\rm e} = n_{\rm p} = n$, we can also substitute the following
\bea
&{\rho_0}_{\rm p} = n m_{\rm p} = n(1.67262158 \times 10^{-27}) \rm{kg}  \\
&{\rho_0}_{\rm e} = n m_{\rm e} = n(9.10938188 \times 10^{-31}) \rm{kg}  \\
&\rho_{\gamma} = \frac{3n(m_{\rm p} + m_{\rm e})}{4R_b} = \frac{n}{R_b}(1.25514939 \times 10^{-27}) \rm{kg}
\eea

\setcounter{section}{2}

\subsection{Constants given in SI units}

\bea
c &=& 2.99792458 \times 10^8 m s^{-1} \nonumber \\
m_p &=& 1.67262158 \times 10^{-27} kg \nonumber \\
\sigma_T &=& 6.65245854533 \times 10^{-29} m^2 \nonumber \\
e &=& 1.60217646 \times 10^{-19} C \nonumber \\
\epsilon_0 &=& 8.854187817620 \times 10^{-12} C^2 kg^{-1} m^{-3} s^2 \nonumber \\
\mu_0 &=& 1.256637 \times 10^{-6} C^{-2} kg m \nonumber \\
\beta &=& 5.446170245 \times 10^{-4} \nonumber \\
Mpc^{-1} &=& 3.24 \times 10^{-23} m^{-1}
\eea

\subsection{Cosmological Parameters}

Values have been taken from Planck results, where these were not available we have used WMAP values. 

\bea
&\eta_0 = \frac{\eta_{eq}}{a_{eq}} = 3.47276 \times 10^{19} s \,,
\hspace{1cm}
&T_b = 2.7255 K \nonumber \\
&\alpha(k_0) = 0.13 \,,
\hspace{1cm}
&\Delta^2_{\R}(k_0) = 2.38 \times 10^{-9} \nonumber \\
&k_0 = 0.002 Mpc^{-1} \,,
\hspace{1cm}
&\rho_c = 9.6594 \times 10^{-27} kg m^{-3} \nonumber \\
&\Omega_0 = 1.02 \,,
\hspace{1cm}
&\rho_{0 \rm{init} } = 9.85 \times 10^{-27} kg m^{-3}\nonumber\\
&z_{\rm{eq}} = 3402\,,
\hspace{1cm}
&a_{\rm{eq}} = 2.94 \times 10^{-4} \nonumber 
\eea

\subsection{Variables}

\bea
n &=& n_e = n_p = n_B = \frac{2\zeta(3)\eta_{B0} }{\pi^2} T^3 = 0.251367 a^{-3} m^{-3}  \equiv \hat{n} a^{-3} m^{-3} \nonumber \\
R_b &=& 698.38 \left( \frac{h_0^2 \Omega_b}{0.022} \right) a \equiv \hat{R}_b a \nonumber \\
\sigma_C &=& \frac{T^{3/2}}{\pi e^2 \sqrt{m_e} \rm{ln}\Lambda} \approx 2 \times 10^8 a^{-3/2} s^{-1}
\eea

\section*{References}
\bibliographystyle{unsrt.bst}
\bibliography{magpaper_18Aug.bbl}

\begin{thebibliography}{10}

\bibitem{Brown:2006wv}
Iain~A. Brown.
\newblock {\em {Primordial Magnetic Fields in Cosmology}}.
\newblock University of Portsmouth, 2006.
\newblock Ph.D. Thesis.

\bibitem{Widrow:2002ud}
Lawrence~M. Widrow.
\newblock {Origin of galactic and extragalactic magnetic fields}.
\newblock {\em Rev.Mod.Phys.}, 74:775--823, 2002.

\bibitem{Kulsrud:2007an}
Russell~M. Kulsrud and Ellen~G. Zweibel.
\newblock {The Origin of Astrophysical Magnetic Fields}.
\newblock {\em Rept.Prog.Phys.}, 71:0046091, 2008.

\bibitem{Kronberg:1993vk}
Philipp~P. Kronberg.
\newblock {Extragalactic magnetic fields}.
\newblock {\em Rept.Prog.Phys.}, 57:325--382, 1994.

\bibitem{Carilli:2001hj}
C.L. Carilli and G.B. Taylor.
\newblock {Cluster magnetic fields}.
\newblock {\em Ann.Rev.Astron.Astrophys.}, 40:319--348, 2002.

\bibitem{Kronberg:2007dy}
P.P. Kronberg, M.L. Bernet, F.~Miniati, S.J. Lilly, M.B. Short, et~al.
\newblock {A Global Probe of Cosmic Magnetic Fields to High Redshifts}.
\newblock {\em Astrophys.J.}, 676:7079, 2008.

\bibitem{Bernet:2008qp}
Martin~L. Bernet, Francesco Miniati, Simon~J. Lilly, Philipp~P. Kronberg, and
  Miroslava Dessauges-Zavadsky.
\newblock {Strong magnetic fields in normal galaxies at high redshifts}.
\newblock {\em Nature}, 454:302--304, 2008.

\bibitem{Wolfe:2008nk}
Arthur~M. Wolfe, Regina~A. Jorgenson, Timothy Robishaw, Carl Heiles, and
  Jason~X. Prochaska.
\newblock {An 84 microGauss Magnetic Field in a Galaxy at Redshift z=0.692}.
\newblock {\em Nature}, 455:638, 2008.

\bibitem{Tavecchio:2010ja}
F.~Tavecchio, G.~Ghisellini, G.~Bonnoli, and L.~Foschini.
\newblock {Extreme TeV blazars and the intergalactic magnetic field}.
\newblock 2010.

\bibitem{Ando:2010rb}
Shin'ichiro Ando and Alexander Kusenko.
\newblock {Evidence for Gamma-Ray Halos Around Active Galactic Nuclei and the
  First Measurement of Intergalactic Magnetic Fields}.
\newblock {\em Astrophys.J.}, 722:L39, 2010.

\bibitem{Neronov:1900zz}
A.~Neronov and I.~Vovk.
\newblock {Evidence for strong extragalactic magnetic fields from Fermi
  observations of TeV blazars}.
\newblock {\em Science}, 328:73--75, 2010.

\bibitem{Tavecchio:2010mk}
F.~Tavecchio, G.~Ghisellini, L.~Foschini, G.~Bonnoli, G.~Ghirlanda, et~al.
\newblock {The intergalactic magnetic field constrained by Fermi/LAT
  observations of the TeV blazar 1ES 0229+200}.
\newblock {\em Mon.Not.Roy.Astron.Soc.}, 406:L70--L74, 2010.

\bibitem{Essey:2010nd}
Warren Essey, Shin'ichiro Ando, and Alexander Kusenko.
\newblock {Determination of intergalactic magnetic fields from gamma ray data}.
\newblock {\em Astropart.Phys.}, 35:135--139, 2011.

\bibitem{moffatbook}
H.~K. {Moffatt}.
\newblock {\em {Magnetic field generation in electrically conducting fluids}}.
\newblock Cambridge, England, Cambridge University Press, 1978.~353 p., 1978.

\bibitem{Kulsrud:1992rk}
Russell~M. Kulsrud and Stephen~W. Anderson.
\newblock {The spectrum of random magnetic fields in the mean field dynamo
  theory of the Galactic magnetic field}.
\newblock {\em Astrophys.J.}, 396:606--630, 1992.

\bibitem{han}
J.~L. {Han}, R.~N. {Manchester}, E.~M. {Berkhuijsen}, and R.~{Beck}.
\newblock {Antisymmetric rotation measures in our Galaxy: evidence for an A0
  dynamo.}
\newblock {\em Astron.Astrophys.}, 322:98--102, June 1997.

\bibitem{Grasso:2000wj}
Dario Grasso and Hector~R. Rubinstein.
\newblock {Magnetic fields in the early universe}.
\newblock {\em Phys.Rept.}, 348:163--266, 2001.

\bibitem{Kulsrud:1996km}
Russell~M. Kulsrud, Renyue Cen, Jeremiah~P. Ostriker, and Dongsu Ryu.
\newblock {The Protogalactic origin for cosmic magnetic fields}.
\newblock {\em Astrophys.J.}, 480:481, 1997.

\bibitem{King:2005xh}
Emma~J. King and Peter Coles.
\newblock {Amplification of primordial magnetic fields by anisotropic
  gravitational collapse}.
\newblock {\em Mon.Not.Roy.Astron.Soc.}, 365:1288--1294, 2006.

\bibitem{biermann}
L.~Biermann.
\newblock Ueber den ursprung der magnetfelder auf sternen und im interstellaren
  raum (miteinem anhang von a. schlueter).
\newblock {\em Z. Naturforsch. Teil {\bf{A}}}, 5:65, 1950.

\bibitem{daly:1990}
R.~A. {Daly} and A.~{Loeb}.
\newblock {A possible origin of galactic magnetic fields}.
\newblock {\em Astrophys. J.}, 364:451--455, December 1990.

\bibitem{subramanian1994thermal}
K~Subramanian, D~Narasimha, and SM~Chitre.
\newblock Thermal generation of cosmological seed magnetic fields in ionization
  fronts.
\newblock {\em Monthly Notices of the Royal Astronomical Society}, 271:L15,
  1994.

\bibitem{Gnedin:2000ax}
Nickolay~Y. Gnedin, Andrea Ferrara, and Ellen~G. Zweibel.
\newblock {Generation of the primordial magnetic fields during cosmological
  reionization}.
\newblock {\em Astrophys.J.}, 539:505--516, 2000.

\bibitem{Davies:1999bk}
George Davies and Lawrence~M. Widrow.
\newblock {The First magnetic fields}.
\newblock 1999.

\bibitem{Giovannini:2003yn}
Massimo Giovannini.
\newblock {The Magnetized universe}.
\newblock {\em Int.J.Mod.Phys.}, D13:391--502, 2004.

\bibitem{Hanayama:2005hd}
Hidekazu Hanayama, Keitaro Takahashi, Kei Kotake, Masamune Oguri, Kiyotomo
  Ichiki, et~al.
\newblock {Biermann mechanism in primordial supernova remnant and seed magnetic
  fields}.
\newblock {\em Astrophys.J.}, 633:941, 2005.

\bibitem{Miranda:1998ne}
Oswaldo~D. Miranda, Merav Opher, and Reuven Opher.
\newblock {Seed magnetic fields generated by primordial supernova explosions}.
\newblock {\em Mon.Not.Roy.Astron.Soc.}, 1998.

\bibitem{Turner:1987bw}
Michael~S. Turner and Lawrence~M. Widrow.
\newblock {Inflation Produced, Large Scale Magnetic Fields}.
\newblock {\em Phys.Rev.}, D37:2743, 1988.

\bibitem{Tornkvist:2000js}
Ola Tornkvist, Anne-Christine Davis, Konstantinos Dimopoulos, and Tomislav
  Prokopec.
\newblock {Large scale primordial magnetic fields from inflation and
  preheating}.
\newblock pages 443--446, 2000.

\bibitem{Dimopoulos:2001wx}
Konstantinos Dimopoulos, T.~Prokopec, O.~Tornkvist, and A.C. Davis.
\newblock {Natural magnetogenesis from inflation}.
\newblock {\em Phys.Rev.}, D65:063505, 2002.

\bibitem{Prokopec:2004au}
Tomislav Prokopec and Ewald Puchwein.
\newblock {Nearly minimal magnetogenesis}.
\newblock {\em Phys.Rev.}, D70:043004, 2004.

\bibitem{Bamba:2004cu}
Kazuharu Bamba and J.~Yokoyama.
\newblock {Large-scale magnetic fields from dilaton inflation in noncommutative
  spacetime}.
\newblock {\em Phys.Rev.}, D70:083508, 2004.

\bibitem{Bassett:2000aw}
Bruce~A. Bassett, Giuseppe Pollifrone, Shinji Tsujikawa, and Fermin Viniegra.
\newblock {Preheating as cosmic magnetic dynamo}.
\newblock {\em Phys.Rev.}, D63:103515, 2001.

\bibitem{Marklund:2000zs}
Mattias Marklund, Peter~K.S. Dunsby, and Gert Brodin.
\newblock {Cosmological electromagnetic fields due to gravitational wave
  perturbations}.
\newblock {\em Phys.Rev.}, D62:101501, 2000.

\bibitem{harrison}
E.~R. Harrison.
\newblock Generation of magnetic fields in the radiation era.
\newblock {\em Mon. Not. R. atr. Soc.}, 147:279, 1970.

\bibitem{Matarrese:2004kq}
S.~Matarrese, S.~Mollerach, A.~Notari, and A.~Riotto.
\newblock {Large-scale magnetic fields from density perturbations}.
\newblock {\em Phys. Rev.}, D71:043502, 2005.

\bibitem{Gopal:2004ut}
Rajesh Gopal and Shiv Sethi.
\newblock {Generation of Magnetic Field in the Pre-recombination Era}.
\newblock {\em Mon. Not. Roy. Astron. Soc.}, 363:521--528, 2005.

\bibitem{Takahashi:2005nd}
Keitaro Takahashi, Kiyotomo Ichiki, Hiroshi Ohno, and Hidekazu Hanayama.
\newblock {Magnetic field generation from cosmological perturbations}.
\newblock {\em Phys. Rev. Lett.}, 95:121301, 2005.

\bibitem{Betschart:2003bn}
Gerold Betschart, Peter~K.S. Dunsby, and Mattias Marklund.
\newblock {Cosmic magnetic fields from velocity perturbations in the early
  universe}.
\newblock {\em Class.Quant.Grav.}, 21:2115--2126, 2004.

\bibitem{Durrer:2013pga}
R.~Durrer and A.~Neronov.
\newblock {Cosmological Magnetic Fields: Their Generation, Evolution and
  Observation}.
\newblock 2013.

\bibitem{Tashiro:2006uv}
Hiroyuki Tashiro and Naoshi Sugiyama.
\newblock {Probing Primordial Magnetic Fields with the 21cm Fluctuations}.
\newblock {\em Mon.Not.Roy.Astron.Soc.}, 372:1060--1068, 2006.

\bibitem{Schleicher:2008hc}
Dominik~R.G. Schleicher, Robi Banerjee, and Ralf~S. Klessen.
\newblock {Influence of primordial magnetic fields on 21 cm emission}.
\newblock {\em Astrophys.J.}, 692:236--245, 2009.

\bibitem{Ichiki:2007hu}
Kiyotomo Ichiki, Keitaro Takahashi, Naoshi Sugiyama, Hidekazu Hanayama, and
  Hiroshi Ohno.
\newblock {Magnetic Field Spectrum at Cosmological Recombination}.
\newblock 2007.

\bibitem{Maeda:2011uq}
Satoshi Maeda, Keitaro Takahashi, and Kiyotomo Ichiki.
\newblock {Primordial magnetic fields generated by the non-adiabatic
  fluctuations at pre-recombination era}.
\newblock {\em JCAP}, 1111:045, 2011.

\bibitem{Fenu:2010kh}
Elisa Fenu, Cyril Pitrou, and Roy Maartens.
\newblock {The seed magnetic field generated during recombination}.
\newblock {\em Mon.Not.Roy.Astron.Soc.}, 414:2354--2366, 2011.

\bibitem{Ryu16052008}
Dongsu Ryu, Hyesung Kang, Jungyeon Cho, and Santabrata Das.
\newblock Turbulence and magnetic fields in the large-scale structure of the
  universe.
\newblock {\em Science}, 320(5878):909--912, 2008.

\bibitem{Barrow:2006ch}
John~D. Barrow, R.~Maartens, and Christos~G. Tsagas.
\newblock {Cosmology with inhomogeneous magnetic fields}.
\newblock {\em Phys.Rept.}, 449:131--171, 2007.

\bibitem{Bardeen:1980kt}
James~M. Bardeen.
\newblock {Gauge Invariant Cosmological Perturbations}.
\newblock {\em Phys. Rev.}, D22:1882--1905, 1980.

\bibitem{ks}
Hideo Kodama and Misao Sasaki.
\newblock {Cosmological Perturbation Theory}.
\newblock {\em Prog. Theor. Phys. Suppl.}, 78:1--166, 1984.

\bibitem{Ade:2013zuv}
P.A.R. Ade et~al.
\newblock {Planck 2013 results. XVI. Cosmological parameters}.
\newblock 2013.

\bibitem{MW2008}
Karim~A. Malik and David Wands.
\newblock {Cosmological perturbations}.
\newblock {\em Phys. Rept.}, 475:1--51, 2009.

\bibitem{thesis}
Adam~J. Christopherson.
\newblock {\em {Applications of Cosmological Perturbation Theory}}.
\newblock PhD thesis, University of London, 2011.

\bibitem{Tsagas:2004kv}
Christos~G. Tsagas.
\newblock {Electromagnetic fields in curved spacetimes}.
\newblock {\em Class.Quant.Grav.}, 22:393--408, 2005.

\bibitem{vorticity}
Adam~J. Christopherson, Karim~A. Malik, and David~R. Matravers.
\newblock {Vorticity generation at second order in cosmological perturbation
  theory}.
\newblock {\em Phys. Rev.}, D79:123523, 2009.

\bibitem{Christopherson:2010dw}
Adam~J. Christopherson and Karim~A. Malik.
\newblock {Can cosmological perturbations produce early universe vorticity?}
\newblock {\em Class.Quant.Grav.}, 28:114004, 2011.

\bibitem{Malik:2002jb}
Karim~A. Malik, David Wands, and Carlo Ungarelli.
\newblock {Large-scale curvature and entropy perturbations for multiple
  interacting fluids}.
\newblock {\em Phys. Rev.}, D67:063516, 2003.

\bibitem{Christopherson:2010ek}
Adam~J. Christopherson, Karim~A. Malik, and David~R. Matravers.
\newblock {Estimating the amount of vorticity generated by cosmological
  perturbations in the early universe}.
\newblock {\em Phys.Rev.}, D83:123512, 2011.

\bibitem{Brown:2011dn}
Iain~A. Brown, Adam~J. Christopherson, and Karim~A. Malik.
\newblock {The magnitude of the non-adiabatic pressure in the cosmic fluid}.
\newblock {\em Mon.Not.Roy.Astron.Soc.}, 423:1411, 2012.

\bibitem{Gordon:2000hv}
Christopher Gordon, David Wands, Bruce~A. Bassett, and Roy Maartens.
\newblock {Adiabatic and entropy perturbations from inflation}.
\newblock {\em Phys. Rev.}, D63:023506, 2001.

\bibitem{seminal}
Karim~A. Malik and David Wands.
\newblock {Adiabatic and entropy perturbations with interacting fluids and
  fields}.
\newblock {\em JCAP}, 0502:007, 2005.

\bibitem{Huang:2012ub}
Zhiqi Huang and Filippo Vernizzi.
\newblock {Cosmic Microwave Background Bispectrum from Recombination}.
\newblock {\em Phys.Rev.Lett.}, 110(10):101303, 2013.

\bibitem{Pettinari:2013he}
Guido~W. Pettinari, Christian Fidler, Robert Crittenden, Kazuya Koyama, and
  David Wands.
\newblock {The intrinsic bispectrum of the Cosmic Microwave Background}.
\newblock {\em JCAP}, 1304:003, 2013.

\bibitem{magneto_nonlinear}
E.~Nalson, A.~J. Christopherson, and K.~A. Malik.
\newblock Revisiting primordial magnetogenesis mechanisms in the presence of
  second order cosmological perturbations.
\newblock (in preparation).

\bibitem{Su:2012gt}
S.-C. Su, Eugene~A. Lim, and E.P.S. Shellard.
\newblock {CMB Bispectrum from Non-linear Effects during Recombination}.
\newblock 2012.

\bibitem{Peeters:2007wn}
Kasper Peeters.
\newblock {Introducing Cadabra: A Symbolic computer algebra system for field
  theory problems}.
\newblock 2007.

\bibitem{Peeters:2006kp}
Kasper Peeters.
\newblock {A Field-theory motivated approach to symbolic computer algebra}.
\newblock {\em Comput.Phys.Commun.}, 176:550--558, 2007.

\end{thebibliography}

\end{document}